\documentclass[10pt,pra,aps,a4paper,oneside,twocolumn,superscriptaddress,showpacs]{revtex4-2}
\usepackage[english]{babel}
\usepackage[latin1]{inputenc}
\usepackage[T1]{fontenc}
\usepackage{lmodern}
\usepackage{ulem}
\usepackage{dcolumn}
\usepackage{subfigure}
\usepackage{footnote}
\usepackage[title]{appendix}                            % puts "appendix" in appendix titles
\usepackage{multirow}
\usepackage{amsmath,amsfonts,amssymb}
\usepackage{graphicx}                                   % for including eps-figure files
\usepackage{latexsym}                                   % useful for coding complex math
\usepackage{color}
\usepackage{xcolor}
\usepackage{soul}
\usepackage{upgreek}                                    % for non-italic greek letters
\usepackage{hyperref}                                   % add hypertext capabilities
\hypersetup{
    colorlinks=true,
    linkcolor={blue},
    citecolor={red},
    urlcolor={blue}
}
\graphicspath{{./Figures/}}

\newcommand{\Eqn}[2]{\begin{equation}\label{#1}#2\end{equation}} 
\newcommand{\Aln}[2]{\begin{align}\label{#1}#2\end{align}}
\newcommand{\<}{\langle}
\renewcommand{\>}{\rangle}

\begin{document}

\title{Quantum resonant optical bistability with a narrow atomic
transition: bistability phase diagram in the bad cavity regime}

\author{\firstname{D.} \surname{Rivero}}
\email{philippe.courteille@ifsc.usp.br}
\affiliation{Instituto de F\'isica de S\~ao Carlos,
Universidade de S\~ao Paulo, S\~ao Carlos, SP 13566-970, Brazil}
\affiliation{Institute for Quantum Electronics, ETH Z\"urich, 8093 Z\"urich, Switzerland}

\author{\firstname{C.A.} \surname{Pessoa Jr.}}
\affiliation{Instituto de F\'isica de S\~ao Carlos,
Universidade de S\~ao Paulo, S\~ao Carlos, SP 13566-970, Brazil}

\author{\firstname{G.H.} \surname{de Fran\c{c}a}}
\affiliation{Instituto de F\'isica de S\~ao Carlos,
Universidade de S\~ao Paulo, S\~ao Carlos, SP 13566-970, Brazil}

\author{\firstname{R.C.} \surname{Teixeira}}
\affiliation{Departamento de F\'isica,
Universidade Federal de S\~ao Carlos, S\~ao Carlos, SP 13565-905, Brazil}

\author{\firstname{S.} \surname{Slama}}
\affiliation{Center for Quantum Science and Physikalisches Institut, Eberhard-Karls Universit\"at T\"ubingen, Auf der Morgenstelle 14, 72076 T\"ubingen, Germany}

\author{\firstname{Ph.W.} \surname{Courteille}}
\affiliation{Instituto de F\'isica de S\~ao Carlos,
Universidade de S\~ao Paulo, S\~ao Carlos, SP 13566-970, Brazil}

\begin{abstract}
We report on the observation of a novel manifestation of saturation-induced optical bistability in a resonantly pumped optical ring cavity interacting strongly with a cloud of atoms via a narrow atomic transition. The bistability emerges, above a critical pump rate, as an additional peak in the cavity's normal mode spectrum close to atomic resonance. This third transmission peak is usually suppressed due to strong resonant absorption, but in our experiment it is visible because of the linewidth of the atomic transition being much smaller than that of the cavity, which sets the experiment into the bad-cavity regime. Relying on complete saturation of the transition, this bistability has a quantum origin and cannot be mimicked by a classical material presenting a nonlinear refraction index. The appearance of the central peak in addition to the normal modes is predicted by a semi-classical model derived from the Tavis-Cummings Hamiltonian from which we derive a bistability phase diagram that connects our observations with former work on optical bistability in the good cavity regime. The phase diagram reveals several so far unexplored bistable phases.
\end{abstract}

\maketitle

%\Dalila{Text inserted by Dalila}
%\Claudio{Text inserted by Claudio}
%\Gustavo{Text inserted by Gustavo}
%\Wastl{Text inserted by Wastl}
%\Raul{Text inserted by Raul}
%\Philippe{Text inserted by Philippe}

Cavity-mediated interactions between atoms provide a promising route to generate non-trivial interatomic correlations leading to phenomena such as spin-squeezing \cite{Kitagawa93,Leroux10,Schleier-Smith10,Schleier-Smith10b,ChenZ11,ChenZ11,Bohnet14,CoxKC16,XuMinghui16,Salvi18}, superradiant lasing \cite{Meiser09,MaierTh14,Norcia16b,Norcia16c,Norcia18b,Davis19,Bychek21,Orioli22}, or quantum state magnification \cite{Hosten16,ClineJ22}, which are of utmost interest for metrological applications. One important request is that the rate of spontaneous emission $\Gamma$ is small, leaving cavity decay occurring at a rate $\kappa$ as the only available decay channel. In this limit, known as the 'bad cavity' regime, non-classical features in the photon statistics become more robust \cite{Lambrecht96,Turchette98,Schnabel16,Louchet-Chauvet09,Dantan06} and quantum correlations more stable. A second requirement is a high collective cooperativity, which guarantees that light emitted from one atom may be reabsorbed by another one and thus information be shared between atoms. These ingredients can lead to nonlinear system dynamics and eventually to bistability. For the optical mode the bistability can manifest as two stable states of transmission, whose realization depend on the history of the system. The bistability can originate in different features of the system. For example in \cite{Gibbs76b,Gripp97,Gripp97b,Gothe19} it is caused by saturation, in \cite{Gupta07,Ritter09} by optomechanical coupling, and in \cite{Lambrecht95,Gabor22,Suarez23} by the multilevel structure of the ground state. Clearly, the bistable region of parameters is the one where to expect quantum correlations \cite{Lambrecht96,Norcia16b}.

Strong interaction between the atoms and the cavity mode leads to normal mode splitting of the cavity's transmission resonance when the frequency of the atomic transition is close to the resonance of the empty cavity \cite{ZhuY90,SchusterI08,MueckeM10,ArnoldKJ11,Ritsch13}. From a semi-classical model of light-atom interaction, the atom-cavity resonances can be found at those frequencies where the acquired phase for a round trip of the light inside the cavity filled with atoms is zero \cite{ZhuY90, Gripp97b}. For high collective cooperativity and moderate intracavity intensities, this zero-phase condition generates three resonances: two that correspond to the normal mode splitting plus a central one that is, in practice, heavily suppressed due to atomic absorption. But the use of a narrow atomic transition can change this scenario and open a saturation-induced transparency window, allowing for an observation of this central resonance.\\ 

\begin{figure*}
\centerline{\includegraphics[width=18truecm]{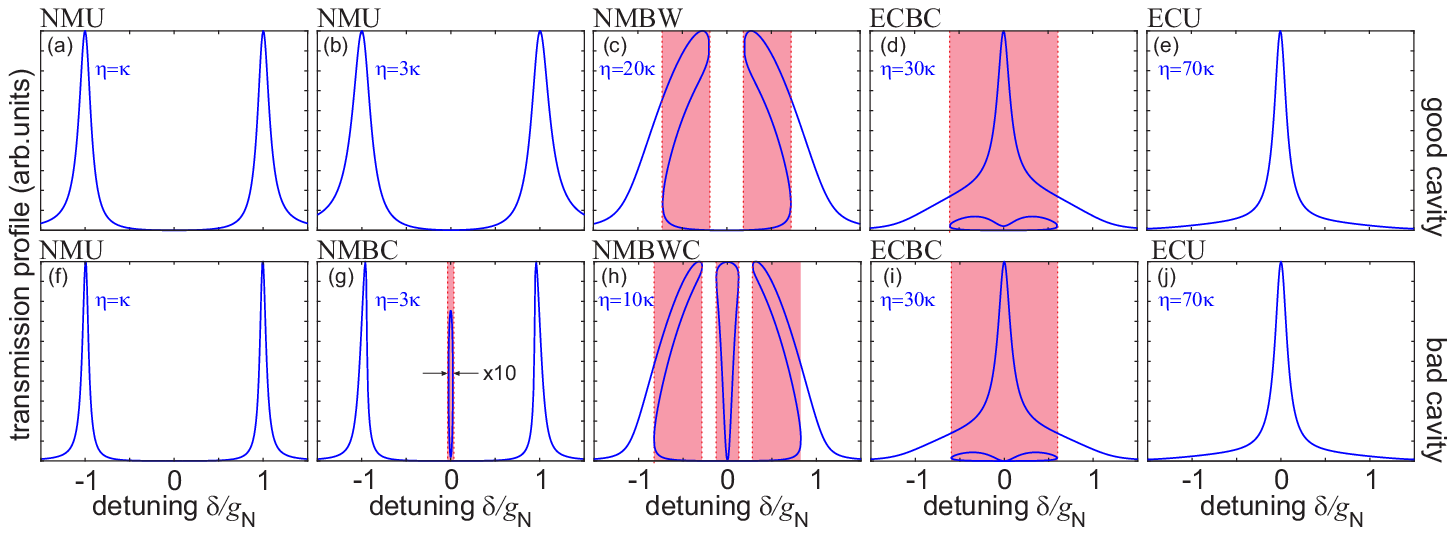}}
    \caption{\small{Saturation-induced phases of strongly coupled atom-cavity systems: Depending on the pumping strength $\eta$ the cavity transmission spectra reveal different unistable and bistable phases (marked in red). These phases are well-known in the good cavity limit (upper row): an increasing pumping strength drives the system from a normal mode unistable phase (NMU) across a normal mode bistable wing phase (NMBW) and an empty cavity-like bistable center phase (ECBC) to an empty cavity-like unistable phase (ECU). The bad cavity limit (lower row) features phases which are so far unexplored: the normal mode bistable center phase (NMBC) and the normal mode bistable wing and center phase (NMBWC). It features also the empty cavity-like bistable center phase (ECBC), appearing in the good cavity limit. The parameters used are $g\sqrt{N}=12.4\kappa$ in all subplots as in \cite{Gripp97}. For the good and bad cavity limit, the atomic linewidth is set to $\Gamma=2\kappa$ and $\Gamma=0.01\kappa$, respectively in the upper and lower row. The pumping strength is chosen as indicated in the subplots in order to exemplarily show the different phases. The central part in subfigure (g) has been horizontally stretched in order to show the narrow feature more clearly.}}
    \label{fig:phases}
\end{figure*}

In this work, contrasting with experiments \cite{Gibbs76b,Lambrecht96,Gripp97,Gothe19} carried out in the good-cavity limit, we choose the narrow intercombination transition $^1S_0\leftrightarrow{^3P}_1$ in atomic strontium with a decay rate $\Gamma/2\pi=7.5\operatorname{kHz}$ for interaction with a cavity mode with amplitude decay width $\kappa/2\pi=3.4\operatorname{MHz}$. This sets the system deep into the bad cavity limit. At the same time the atom-field coupling strength $g_N=g\sqrt{N}$ with coupling constant $g/2\pi=9.1\operatorname{kHz}$ is collectively enhanced by a number of up to $N=200\,000$ atoms such that the atoms are strongly coupled to the cavity. This allows us to probe the normal modes in the bad cavity limit featuring a series of bistable phases, which are different from the ones in the good cavity limit. Fig.~\ref{fig:phases} shows normal mode spectra calculated for different cavity pump rates $\eta$. In particular, we can excite and study the resonant normal mode, which is suppressed in the good cavity limit \cite{Gripp97}. In Fig.~\ref{fig:phases} this is seen by comparing the spectra (c) and (h). Note that the used transition has a low enough saturation intensity to be saturated with even single photons in our cavity.\\

In Sec.~\ref{sec:Experiment} we present our experimental setup, procedure, and measurements. In Sec.~\ref{sec:Theory} we expose our model derived from the many-atom Dicke Hamiltonian. The stationary intracavity photon number predicted by the model allows us to calculate the normal modes under special consideration of the respective roles of atomic bunching, particular pumping schemes of the ring cavity modes, and saturation of the atomic transition. We verified that our theoretical model describes the instabilities observed in other experiments \cite{Gripp97b,Gothe19}. The measurements are quantitatively compared with our model and interpreted in Sec.~\ref{sec:Discussion}.

\section{Experiment} \label{sec:Experiment}

\subsection{Experimental setup} \label{sec:ExperimentSetup}

\begin{figure*}[ht]
    \centerline{\includegraphics[width=18truecm]{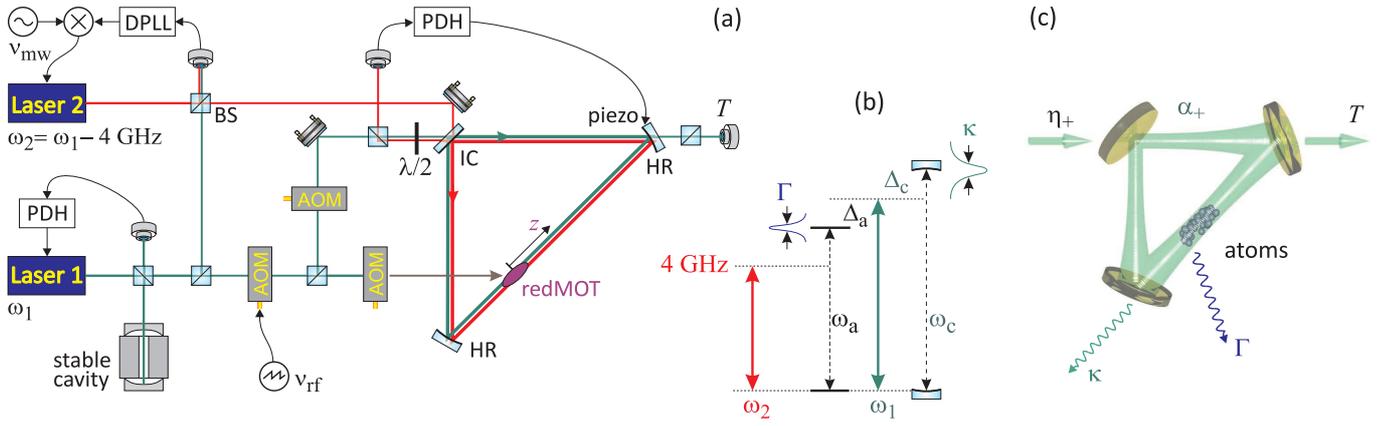}}
    \caption{\small{(a)~Scheme of the lasers controlling the atom-cavity interaction: Laser~1, which is tightly locked to a reference cavity via a Pound-Drever-Hall servo (PDH), has a small detuning $\Delta_\mathrm{a}$ from the narrow atomic resonance. It serves for cooling the atomic cloud (red MOT) and for driving one of the counter-propagating modes ($\alpha_+$) of the ring cavity. The  ring cavity is formed by three mirrors, the input coupler (IC) and two high reflectors (HR). The transmission $T$ through a high reflectors is recorded. A second laser (laser~2) is phase-locked (DPLL) to laser~1 at $4\operatorname{GHz}$ below the atomic resonance. In contrast to laser~1, its intensity is high. Coupled into one of the counter-propagating modes of the ring cavity it generates an optical dipole potential in which the atomic cloud is confined. Transmission spectra showing normal mode splitting are obtained by ramping $\nu_\text{rf}$ and/or $\nu_\text{mw}$.
    (b)~Detunings of the laser with respect to the atomic transition and to the ring cavity.
    (c)~Simplified scheme of the ring cavity illustrating dissipation processes due to spontaneous decay ($\Gamma$) and cavity transmission ($\kappa$).}}
    \label{fig:FigLaserScheme}
\end{figure*}
The experimental setup has been presented in detail in Ref.~\cite{Rivero22}. In short, a cloud of about $N=200\,000$ $^{88}$Sr atoms is cooled to temperatures around $1\operatorname{\upmu K}$ and transferred into the mode volume of a ring cavity operated near the narrow intercombination line with the transition wavelength $\lambda_\mathrm{a}=689\operatorname{nm}$ connecting the energy levels $(5s)^2~^1S_0$ and $(5s5p)~^3P_1$. The atomic cloud is confined to the mode volume by the optical dipole potential formed by laser beams coupled into a TEM$_{00}$ mode of the ring cavity (laser~2 in Fig.~\ref{fig:FigLaserScheme}) under $p$-polarization, where the finesse of the cavity is low ($F\approx 100$). The length of the ring cavity is stabilized to this laser~2 via a Pound-Drever-Hall servo electronics. Laser~2, which is tuned $\Delta_\text{dip}/2\pi\approx -530\Gamma$ to the red of the strontium resonance is, in turn, locked via a digital phase-locked loop (DPLL) to a second laser (laser~1 in Fig.~\ref{fig:FigLaserScheme}), which itself is stabilized to a stable supercavity and operated near the strontium resonance. 

Laser~1 serves for the last cooling stage of the atomic cloud, and also as a probe for the atom-cavity interaction. In this latter role, the light of laser~1 is passed through AOMs, driven by a tunable radio-frequency $\nu_\text{rf}$ and permitting to ramp the laser frequency $\omega$ with respect to the atomic resonance $\omega_\mathrm{a}$, thus generating a controlled detuning $\Delta_\mathrm{a}=\omega-\omega_\mathrm{a}$, and with respect to the ring cavity resonance $\omega_\mathrm{c}$, generating a detuning $\Delta_\mathrm{c}=\omega-\omega_\mathrm{c}$. On the other hand, we can control the detuning between the cavity and the atomic resonance, $\Delta_\mathrm{ca}=\omega_\mathrm{c}-\omega_\mathrm{a}=\Delta_\mathrm{a}-\Delta_\mathrm{c}$, via the microwave-frequency $\nu_\mathrm{mw}$ of the DPLL \cite{Rivero22}.

Injected into one of the two counter-propagating modes of the ring cavity, laser~2 generates an optical dipole trap, in which the atomic cloud adopts a smooth density distribution elongated along the cavity's optical axis. Part of the light power of laser~1 is coupled as probe light into a TEM$_{00}$ mode of the ring cavity using $s$-polarization, where the finesse of the cavity is high ($F\approx 1200$), counter-propagating with laser~2. The probe light transmitted through the pumped mode through a high reflecting mirror (transmission $T_\mathrm{hr}\approx 0.001$) is filtered by a polarizing beamsplitter, coupled into optical fibers and monitored. We infer the intracavity power $P$ from the transmitted probe light power
$P_\mathrm{hr}=T_\mathrm{hr}P=T_\mathrm{hr}\hbar\omega\delta_\mathrm{fsr}n$, where $\delta_\mathrm{fsr}=8.23\operatorname{GHz}$ is the free spectral range of the ring cavity and $n$ the intracavity photon number. Subscripts $\pm$ refer to counter-propagating modes of the ring cavity. In the following we assume that only the mode denoted by $\alpha_+$, where $|\alpha_+|^2=n$, is pumped and that the mode $\alpha_-$ is filled only via photonic backscattering by the atomic cloud.

The cavity pump rate $\eta_+$ is an experimental parameter which can be determined from the photon number in the empty cavity under resonant driving, $\eta_+/\kappa=\alpha_\eta\equiv\alpha_+(N=0,\Delta_\mathrm{c}=0)$. It allows to express the ring cavity transmission as, $T=|\alpha_+/\alpha_\eta|^2$. With the saturation parameter given by $s\equiv 2\Omega^2/\Gamma^2$ defined via the Rabi frequency $\Omega\equiv 2g|\alpha_+|$ we find that the transition is saturated by a mean photon number of $|\alpha_\mathrm{sat}|^2\approx 0.087$ corresponding to a power of $P_\mathrm{sat}=\hbar\omega\delta_\mathrm{fsr}|\alpha_\mathrm{sat}|^2 \approx 0.2\operatorname{nW}$. In practice, we typically use saturation parameters in the range $s=P/P_\text{sat}\approx 100...1000$. The single-photon saturation $s_1$ and the single-atom cooperativity $\Upsilon$ are important parameters characterizing our coupled atom-cavity system,\Eqn
    {eq:01}{s_1 \equiv \frac{8g^2}{\Gamma^2} \approx 11.5 ~~~~~,~~~~~ 
        \Upsilon \equiv \frac{4g^2}{\kappa\Gamma} \approx 0.013~.}

\subsection{Measurements} \label{sec:ExperimentMeasure}

\begin{figure}[ht]
    \centerline{\includegraphics[width=9.0truecm]{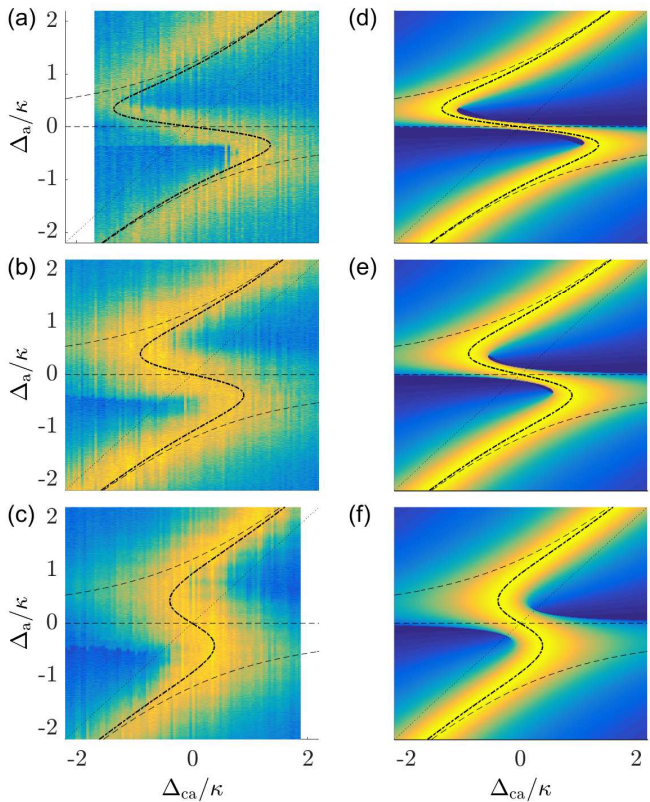}}
    \caption{\small{(a-c)~Measured cavity transmission for various $\eta_+$ taken with $N=200\,000$ atoms: The pump rate is varied from (a)~$\eta_+=110\kappa$ over (b)~$140\kappa$ to (c)~$200\kappa$. 
    (d-f)~Calculated transmission spectra using Eq.~\eqref{eq:03} for pump rates corresponding to those of (a-c) and all other experimental parameters as specified in the text. For the spectral regions presenting bistability, the solution shown is the one with highest transmission. The diagonal black dotted lines show the mode of an empty cavity ($N=0$). The black dash-dotted lines are calculated from Eq.~\eqref{eq:04} using a saturation broadening estimated from $\Omega_\eta=2g\eta_+/\kappa$. The dashed lines correspond to setting $\Omega_\eta=0$.}}
    \label{fig:FigAvoidedCrossing}
\end{figure}

We record normal mode spectra by scanning $\Delta_\mathrm{a}$ for fixed $\Delta_\mathrm{ca}$. The typical duration of a scan is $\Delta t_\mathrm{scan}=0.25\operatorname{ms}$. Repeating the scan for various $\Delta_\mathrm{ca}$ we obtain the two-dimensional spectra shown in Figs.~\ref{fig:FigAvoidedCrossing}(a-c) with false-color coding. Yellow indicates maximum cavity transmission and blue an impermeable cavity reflecting the pump light completely.

The Figs.~\ref{fig:FigAvoidedCrossing}(a-c) show transmission spectra measured for different pump rates $\eta_+$. The spectra show a number of interesting features: (i)~The normal modes are clearly visible as two resonances describing an avoided crossing. These resonances are emphasized by dashed black lines in the figures. (ii)~Near the atomic resonance, around $\Delta_\mathrm{a}\simeq 0$, the spectra develop a narrow ridge of high transmission bridging the normal modes. Normal mode spectra recorded by scanning $\Delta_\mathrm{a}$ near the atom-cavity resonance, i.e.~around $\Delta_\mathrm{ca}\simeq 0$, therefore exhibit three peaks instead of two. (iii)~The ridge is flanked by discontinuous steep precipices into dark regions appearing as sharp horizontal features in the false color plots. The additional central peak has sharp edges, pointing to bistable behavior of the cavity transmission.

Comparing different spectra taken for different pump rates $\eta_+$ we notice additional features: (iv)~The (vertical) width of the bridge between the normal modes in Figs.~\ref{fig:FigAvoidedCrossing}(a-c), that is, the width of the central peak in $\Delta_\mathrm{a}$-scans, increases with stronger pump rates $\eta_+$. (v)~As the pump rate is increased, the discontinuous ledges are 'pushed' away from resonance towards higher $|\Delta_\mathrm{ca}|$. At sufficiently strong pumping the avoided crossing of the normal modes turns into a dispersive lineshape. $\Delta_\mathrm{a}$-scans near atom-cavity resonance still show three peaks, but do not any longer exhibit discontinuities arising from bistable behavior.

\begin{figure}[ht]
\centerline{\includegraphics[width=9.0truecm]{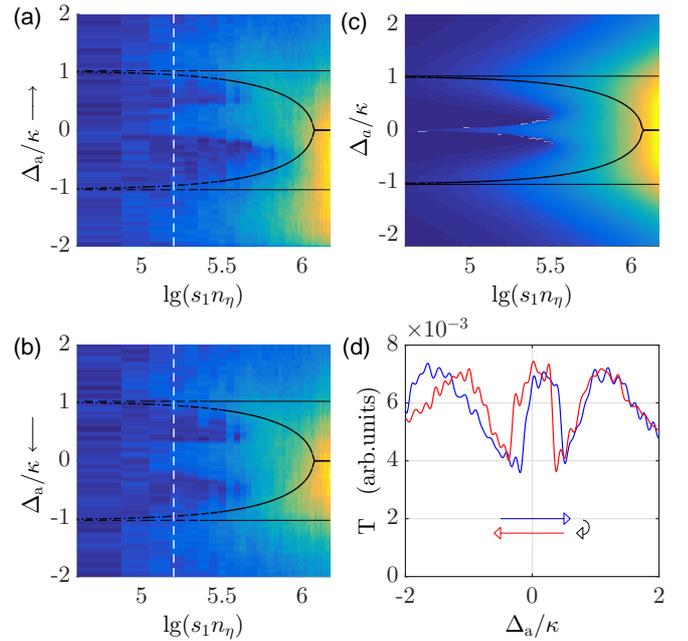}}
    \caption{\small{Pump rate dependence of the normal mode spectrum: (a) resp.~(b) show normal mode spectra taken on resonance ($\Delta_\mathrm{ca}=0$) as a function of pump laser power, $s_\eta=s_1n_\eta=s_1\eta_+^2/\kappa^2$, for the atom-cavity detuning scanned in real time from lower to higher (resp.~higher to lower) values. Above saturation the distance between the normal modes diminishes with increasing pump power. (c)~Simulation based on the same equations as for Fig.~\ref{fig:FigAvoidedCrossing} using the same experimental parameters as for curves (a) and (b). The dashed lines show the low saturation value of the normal mode splitting. The dash-dotted lines denote the frequencies of the normal modes calculated from Eq.~\eqref{eq:04}. The measurements also show a hysteresis-like behavior. Looking at vertical cuts along the white dashed line traced in Figs.(a-b) one observes that the jump between opaque and transparent solutions is always delayed in the scanned direction (blue and red curves in Fig.~(d), the arrows indicate the scan direction). Note that the scan back sees the second normal mode (left peak of red curve) at reduced atomic distance. We attribute this to heating or trap loss during the scan causing a reduction of the normal mode splitting.}}
    \label{fig:FigIntensityScan}
\end{figure}

Fig.~\ref{fig:FigIntensityScan} shows a more detailed study of the dependence of the normal mode spectra on the pump rate $\eta_+$. In Fig.~\ref{fig:FigIntensityScan}(a) the pump laser was scanned from $\Delta_a=-2.2\kappa$ to $+2.2\kappa$, immediately followed by a scan in the inverse direction exhibited in (b). The features to be extracted from these measurements are: (vi)~There is a central peak around $\Delta_\mathrm{a}=0$ bordered by sharp edges. The central peak corresponds to the 'bridge' of Figs.~\ref{fig:FigAvoidedCrossing}(a-c). (vii)~We note a hysteretic behavior of the sharp edges, as the position of the central peak in Fig.~\ref{fig:FigIntensityScan}(b) is slightly shifted as compared to (a). The latter feature is seen more clearly in a single scan taken at a pump rate of $\eta_+=120$ [white dashed lines in Figs.~\ref{fig:FigIntensityScan}(a-b)] and plotted in Fig.~\ref{fig:FigIntensityScan}(d). (viii)~The central peak increases its width at high pump rates. (ix)~The distance between the normal mode peaks diminishes when increasing the pump rate very far beyond saturation. Eventually all peaks merge into a single one. 

We identify in all those features a bistability mechanism of a different kind than previously reported in the literature for atom-cavity systems. In the next sections we develop a semiclassical theory that fits well our data and grasps this new mechanism~\footnote
    {Note that the decrease of the normal mode splitting at high intensities seen in various figures [e.g.~black lines in Fig.~4(c) and Fig.~5(b)] is a consequence of the nonlinearity of the system even in regimes where no bistability is observed.}.

\section{Theoretical description} \label{sec:Theory}

The energy spectrum of the coupled Tavis-Cummings system consisting of $N$ atoms and one cavity mode has been calculated in Refs.~\cite{Narducci73,Garraway11}. In the following we will restrict to calculating the transmission spectra  $T(\Delta_\mathrm{ca},\Delta_\mathrm{a})$ for our ring cavity \footnote{Note that all features observed with a ring cavity in this work also apply to linear cavities.}.

The relevant degrees of freedom of our system are the counter-propagating field modes of the ring cavity, described by the field operators $\hat a_\pm$, and the excitation of atoms located at positions $z_j$ along the cavity's optical axis, denoted by the Pauli operators $\hat\sigma_j^\pm$ and $\hat\sigma_j^z$. To obtain an expression describing normal mode splitting of our coupled atom-ring cavity system (see Appendix \ref{sec:AppDerivation}) we write down the collective Hamiltonian and derive Heisenberg equations for the degrees of freedom. Spontaneous emission and cavity decay are accounted for by jump operators for the dissipative processes occurring at rates $\Gamma$, respectively, $\kappa$. Neglecting all quantum correlations we perform a semiclassical decorrelation \cite{Gripp97} and derive the steady-state solution for the expectation values $\alpha_\pm=\<\hat a_\pm\>$. It reads (see Eq.~\eqref{eq:A16} of the Appendix),\Eqn
    {eq:02}{\sum_{j=1}^N\frac{U_\gamma(\alpha_\pm+e^{\mp 2\imath kz_j})\alpha_\mp}
        {1+2|U_\gamma/g|^2|e^{\imath kz_j}\alpha_++e^{-\imath kz_j}
        \alpha _-|^2} = \imath\eta_\pm-\Delta_\kappa\alpha_\pm~.}
with the abbreviations, $U_\gamma\equiv U_0-\imath\gamma_0\equiv g^2/(\Delta_\mathrm{a}+\imath\tfrac{\Gamma}{2})$ and $\Delta_\kappa\equiv\Delta_\mathrm{c}+\imath\kappa$.

The expression~\eqref{eq:02} can not be solved analytically except in some particular limits. The first simplification consists in assuming that the cavity is pumped from one side only, $\eta_-=0$, and that the cloud is homogeneously distributed over the mode volume of the cavity. In that case, we may neglect backscattered light altogether, $\alpha_-=0$, and Eq.~\eqref{eq:02} simplifies to,\Eqn
    {eq:03}{\alpha_+ = \frac{\imath\eta_+}{\Delta_\kappa-\frac{NU_\gamma}
        {1+2|U_\gamma\alpha_+/g|^2}}~.}
We note that this expression coincides with the equation~(10) used by Gripp et al.~\cite{Gripp97}.

Taking the modulus squared of this expression, we arrive to a cubic equation for the average number of photons inside the cavity mode $n=|\alpha_+|^2$ [Eq.~\eqref{eq:A23} of the Appendix], which presents more than one stable solution for a certain range of parameters. This bistable behavior leaves an experimental signature in the form of sudden jumps of the cavity transmission upon changing certain parameters, e.g.~the pump light intensity or detuning, as already mentioned in the discussion of Figs.~\ref{fig:FigAvoidedCrossing}(a-c) in Sec.~\ref{sec:ExperimentMeasure}. The theoretical transmission spectra exhibited in Figs.~\ref{fig:FigAvoidedCrossing}(d-f) are proportional to the intracavity photon number $n$ and calculated for the same experimental parameters as Figs.~\ref{fig:FigAvoidedCrossing}(a-c), respectively. For the bistable  parameter regions, where more than one stable solution exists, the solution with highest transmission is shown.

\subsection{On-resonance and off-resonance bistability}

The condition of zero phase shift for the intracavity light field $\alpha_+$ with respect to the pump field $\eta_+$ can be found by setting the imaginary part of the right-hand side of Eq.~\eqref{eq:03} to zero. It yields,\Eqn
    {eq:04}{\Delta_\mathrm{ca} = \Delta_\mathrm{a}-\frac{Ng^2\Delta_\mathrm{a}}{\Delta_\mathrm{a}^2+\Gamma^2/4+\Omega_\eta^2/2}~.}
In this expression $\Omega_\eta=2g\eta_+/\kappa$ is the maximum Rabi frequency, i.e.~the Rabi frequency if all impinging photons would enter the cavity. The corresponding curve is shown in Fig.~\ref{fig:FigAvoidedCrossing} as dash-dotted lines. We also show in dashed lines the corresponding zero-phase shift curve for vanishing pump field, $\Omega_\eta\rightarrow 0$. We note that it follows closely the maxima of the experimental transmission curves for all range of experimental parameters. Although this condition is generic for most experiments dealing with cavities filled with atoms and operated in the strong collective coupling regime, the central peak is commonly suppressed due to strong resonant absorption of the light by the atoms. In our case, the width of the atomic transition is very small, and we can expect that very little light intensity will already saturate the transition, thus changing this scenario. The rate of photon injection into the cavity, $\eta_+=\kappa n_\eta$, that saturates all atoms roughly equalizes the rate at which they emit photons once they are saturated, $N\Gamma/2$. This means that for $n_\eta\gtrsim N\Gamma/2\kappa$, the atoms become transparent, and a transmission window opens around the empty cavity resonance. This quantum nonlinear feature of the cavity transmission entails the appearance of a bistable spectral region near resonance. The calculations in the Appendix [see \eqref{eq:A54}] indeed show that for $\Delta_\mathrm{a}=\Delta_\mathrm{c}=0$ a lower bound on the pump rate for the appearance of two stable solutions (the signature of a bistable behavior) is,\Eqn
    {eq:05}{n_\eta \gtrsim \frac{N\Gamma(1+\sqrt{3}/2)}{2\kappa}~,}
and the width of the bistable region for $\Delta_\mathrm{c}=0$ is given by Eq.~\eqref{eq:A62} as,\Eqn
    {eq:06}{|\Delta_\mathrm{a}| \simeq \frac{g}{\sqrt{2}}\sqrt{n_\eta-\frac{N\Gamma}{2\kappa}} 
    = \frac{1}{2} \sqrt{\frac{\Omega_\eta^2}{2}-\frac{\Gamma^2\Upsilon_N}{4}}~.}
We will call this bistability mechanism \textit{on-resonance bistability}. It is present for example in the NMBC phase shown in Fig.~\ref{fig:phases}(g).

Gripp et al.~\cite{Gripp97} find a qualitatively different modification of the normal mode splitting when the intensity is increased, reproduced here in Figs.~\ref{fig:phases}(a-e). For their parameters, the bistability emerges off-resonance, at the inner edge of each of the normal mode splitting peaks. It appears for example in the NMBW phase shown in Fig.~\ref{fig:phases}(c). The underlying mechanism follows a prototypical intensity-driven bistability also found in anharmonic classical mechanical oscillators \cite{Landau76} subject to an intensity-dependent frequency shift of the normal mode splitting resonances. Upon further increase of the pump intensity, this bistability mechanism eventually leads to a merger of the two normal mode peaks into a single transmission peak. For $\kappa,\Gamma\ll g\sqrt{N}$ (i.e., for a normal mode splitting much larger than the width of the normal mode resonances), this happens when the two local extremes of the zero-phase condition curve $\Delta_\mathrm{ca}$ as a function of $\Delta_\mathrm{a}$ as given by Eq.~\eqref{eq:04} merge into a single  point ($\left.d\Delta_\mathrm{ca}/d\Delta_\mathrm{a}\right|_{\Delta_\mathrm{a}=0}=0$). This is the case when\Eqn
    {eq:07}{Ng^2 = \Gamma^2/4+\Omega_\eta^2/2~.}
    
We may reasonably expect that, in order to have on-resonance bistability, the critical intensity for its appearance, as described by Eq.~\eqref{eq:05}, must be smaller than the critical intensity for the merging of the off-resonance bistable spectral intervals. For $Ng^2\gg\Gamma^2/4$, this amounts to $\Gamma(1+\sqrt{3}/2)\lesssim\kappa$, which is qualitatively the condition for the bad-cavity regime. Thus, it is the narrowness of the strontium transition that allows us to observe the new bistable mechanism in our experiment. This new mechanism for bistability relies not only on the nonlinearity of the atom-cavity interaction, but also on the complete saturation of the atomic population to cause full transparency of the cavity to light. Thus, it cannot be generated by a simple nonlinear susceptibility.

\subsection{Comparison to experiment}

The calculated spectra in Figs.~\ref{fig:FigAvoidedCrossing} and  Fig.~\ref{fig:FigIntensityScan} are in good qualitative agreement with the measured ones. For instance, comparing the spectra of Figs.~\ref{fig:FigIntensityScan}(a) and (b) we notice that the bistability around the central peak is symmetrically inverted when the scan direction is also inverted, and that the maximum of the transmission is roughly independent of the pump rate, showing no anharmonicity in the resonance frequency. Those features are also evident in the theoretical spectra of Figs.~\ref{fig:phases}(f-j).

On the other hand, while the observed spectral width of the normal mode resonances agrees well with the simulations and does not depend on the pump rate, the spectral width of the bistable region is systematically larger than predicted by simulations. As illustrated in Fig.~\ref{fig:FigIntensityScan}(c), the width of the bistable region should be dominated by power broadening. But from earlier work \cite{Lambrecht95,Gripp97,Gothe19} we know that the onset of bistability is accompanied by oscillations and hysteresis, that can lead to deformations and broadening of the spectra not grasped by our \textit{stationary model}.

As mentioned in Sec.~\ref{sec:ExperimentMeasure}, we indeed observe hysteresis behavior. The oscillations are manifestations of a ringing induced by a time lag between the evolution imposed by the scan and the evolution of the degrees of freedom characterizing the system, which becomes relevant when the slowest time constant of our system, which is the atomic decay rate $2\pi/\Gamma=0.13\operatorname{ms}$, is comparable to or slower than the scan duration $\Delta t_\mathrm{scan}=0.25\operatorname{ms}$. We indeed observe deformation of the spectra when shorter scan times $\Delta t_\mathrm{scan}$ are chosen. However, our photodetector, being optimized for high sensitivity to weak light intensities, is too slow to resolve the fast oscillations.

An example for a spectral deformation can be observed in Fig.~\ref{fig:FigIntensityScan}(d), where the normal mode splitting is found to be larger on the first scan from negative to positive detunings (blue curve) than on the subsequent scan in backward direction (red curve). The deformations of the spectra become stronger when \textit{longer} scan times are used. A possible reason can be heating of the atomic cloud during the scan causing a reduction of the effective number of atoms interacting with the cavity.

%Another possible source of inhomogeneous broadening effects can be the fact that atoms at different locations of the optical dipole trap are subject to different Stark-shifts with respect to the atomic resonance. Doppler-broadening is negligible at the ultralow temperature of the atomic cloud. However, the fact that only the bistable region is broadened seems to exclude inhomogeneous broadening effects, which should affect all resonances.

\section{Discussion} \label{sec:Discussion}

At vanishing pump rate, the shape of the transmission spectrum of the cavity strongly interacting with the atomic cloud depends only on the collective cooperativity $\Upsilon_N$. In particular, for $\Upsilon_N\gtrsim 1$ the spectrum presents a splitting of the resonance reminiscent of two strongly coupled harmonic oscillators.

For $\Upsilon_N \gtrsim 1$ and a non-vanishing pump rate, the nonlinearity of the atom-light interaction changes this simple scenario. Two other parameters then together determine whether the system will be bistable or not and in which frequency interval bistability will occur. The first parameter is, obviously, the pump rate $\eta_+$ which can be represented either by the average photon number $n_\eta$ that it generates inside the cavity at its bare resonance, or by the saturation parameter $s_\eta=s_1n_\eta$.

\begin{figure}[ht]
    \centerline{\includegraphics[width=9.0truecm]{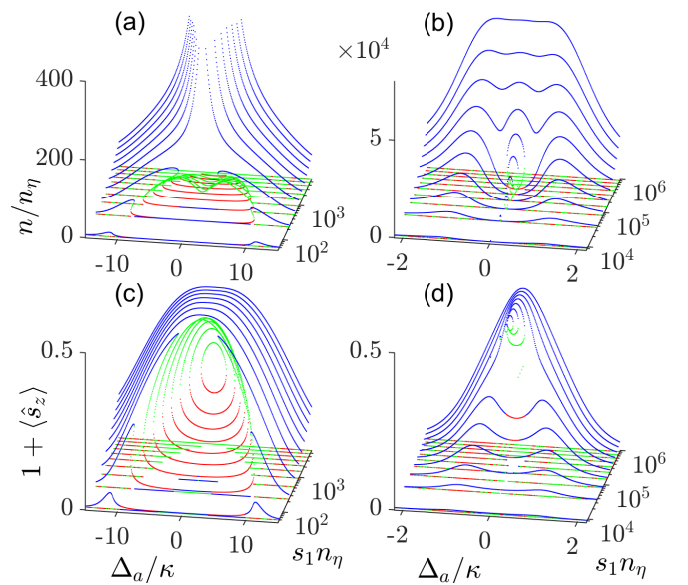}}
    \caption{\small{(a,b) Transmission profiles as a function of $\Delta_\mathrm{a}$ and $\eta_+$ obtained from the analytical solution of Eq.~\eqref{eq:03}. The atomic decay width and the collective atom-cavity coupling strength were chosen for (a)~$\Gamma=2\kappa$ and $g_N=12.4\kappa$ and for (b)~$\Gamma=0.0022\kappa$ and $g_N=1.2\kappa$. Different colors are used to highlight different solutions in the bistable regime.
    (c,d)~Spectra of the excited state populations corresponding to (a,b) calculated from the expression \eqref{eq:A30} in the Appendix.}}
    \label{fig:FigBistableBifurcation}
\end{figure}

Another crucial parameter for the emergence of bistability is the ratio between spontaneous and cavity decay width, $\Gamma/\kappa$, which is the parameter that distinguishes between the good and bad cavity regime. In Fig.~\ref{fig:FigBistableBifurcation} we calculate the transmission profile as a function of $\Delta_\mathrm{a}$ and $\eta_+$ obtained from the analytical solution of Eq.~\eqref{eq:03}. In Figs.~\ref{fig:FigBistableBifurcation}(b,d) the atomic decay width and the atom-cavity coupling strength were chosen as those of our experiment, $\Gamma=0.0022\kappa$ and $g_N=1.2\kappa$. The emergence of on-resonance bistability is clearly visible in a wide range of pump rates. On the other hand, the spectra exhibited in Figs.~\ref{fig:FigBistableBifurcation}(a,c) were calculated for a much stronger spontaneous decay rate $\Gamma=2\kappa$ and $g_N=12.4\kappa$, corresponding to the parameters used by Gripp et al.~\cite{Gripp97}, and show a bistable behavior that emerges from the deformation of the normal mode peaks, until they merge to a single peak for high enough saturation.

\subsection{Bistability phase diagram}

\begin{figure*}[ht]
    \includegraphics[width=18truecm]{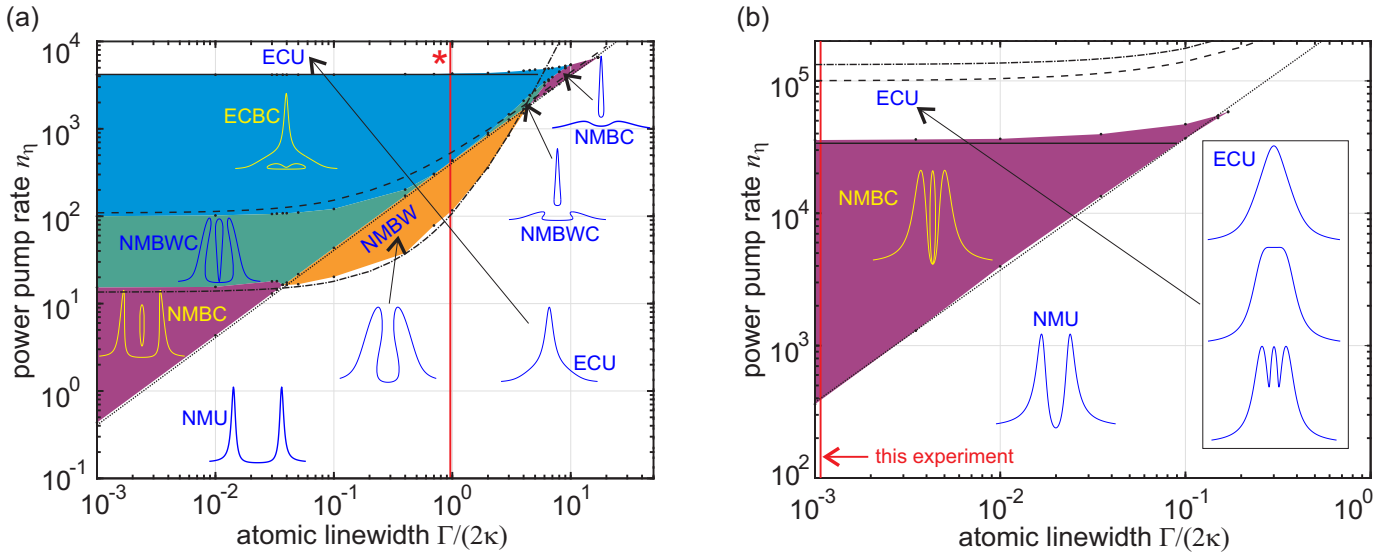}
    \caption{\small{Simulated phase diagram of the coupled atom-cavity system with saturation for (a) collective coupling strength $g_N=12.4\kappa$ as in \cite{Gripp97} (marked by \textcolor{red}{$\ast$}) and (b) $g_N=1.2\kappa$ as in our experiment. Observation of the NMBC and NMBWC phase is featured within the bad cavity limit, while the NMBW phase disappears in the bad cavity limit. The black lines separating the different phases are given by analytical formulas. The dotted diagonal line marks the onset of on-resonance bistability, and the dash-dotted line that at the wings. The dashed line indicates the merging of bistability at the wings and at the center, and finally the solid line marks the maximum pump rate to show bistability.
    Note that NMBWC- and NMBC-like phases exist also in the strongly coupled good cavity limit, i.e. for $\kappa<\Gamma/2<g_N$. In the case of weak coupling reached for even larger atomic linewidth no bistable phase exists. The phase diagram (b) containing the experimental situation is much simpler because of the weaker coupling strength $g_N=1.2\kappa$. Here, only the NMBC bistable phase survives.}}
    \label{fig:phase_diagram_both}
\end{figure*}
In order to derive the complete phase diagram we have solved the steady-state equations for the transmission spectrum for various values of the atomic transverse decay rate $\Gamma/2$ and the pump rate $\eta_+$, both scaled to the cavity linewidth $\kappa$. Some examples have already been plotted in Fig.~\ref{fig:phases}. Details on how we solve the equations are given in the Appendix \ref{sec:AppSolution}. 

We classify the bistable solutions by the overall shape of the spectrum, i.e.~resembling either a normal mode or an empty cavity spectrum, and by the frequencies at which the bistability occurs, i.e.~either in the wings or in the center of the spectrum. In Figs.~\ref{fig:phase_diagram_both}(a) and (b) we compiled phase diagrams for two different situations: Fig.~\ref{fig:phase_diagram_both}(a) is done for $g_N=12.4\kappa$, reproducing the situation of Gripp et al. \cite{Gripp97}, while Fig.~\ref{fig:phase_diagram_both}(b) is obtained for the collective coupling constant $g_N$ of our experimental system.

\begin{figure*}[ht]
   \includegraphics[width=18truecm]{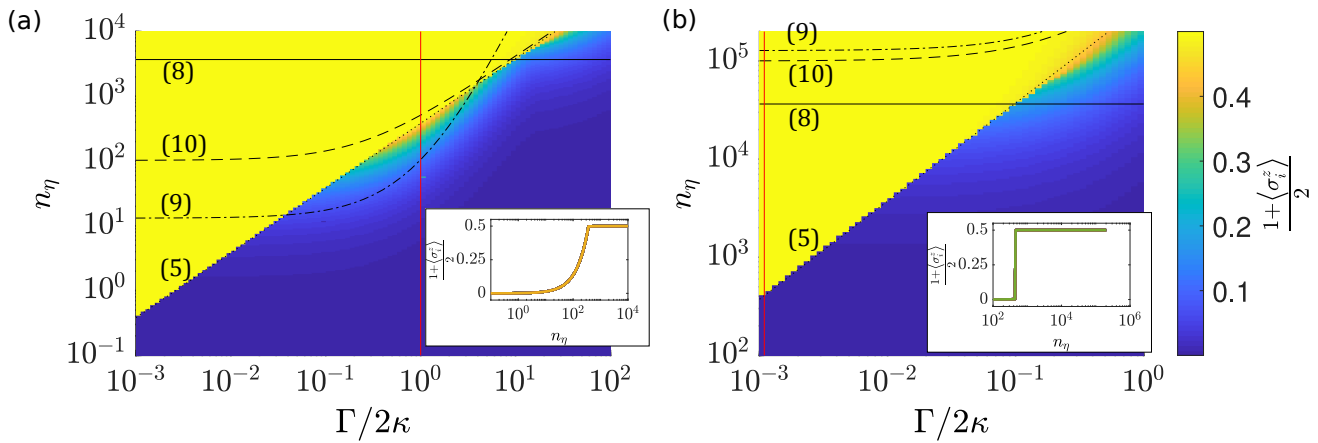}
   \caption{\small{(a) (resp. (b)): Maximum of the atomic excited state population $(1+\<\sigma_1^z\>)/2$ for all light detunings, as a function of $n_\eta$ and $\Gamma/2\kappa$, for same parameters as in Fig.~\ref{fig:phase_diagram_both}(a) (resp.~(b)). The continuous, dashed, dot-dashed and dotted black lines represents the limits between the different bistable regions, identical to the limits in Fig.~\ref{fig:phase_diagram_both}, and are discussed at the main text; The numbers in parentheses correspond to the equations at the main text defining the limits. Inset of (a): Atomic population as a function of $n_\eta$ for a scan of $n_\eta$ following the red vertical line of Fig.~(a), showing a continuous evolution of the atomic population when one goes form a region of no bistability to a region of off-resonance bistability. Inset of (b): Atomic population as a function of $n_\eta$, for a scan of $n_\eta$ following the red vertical line of Fig.~(b), showing a discontinuity at the atomic population when going from a region of no bistability to a region of on-resonance bistability.}}
    \label{fig:population_both}
\end{figure*}

The parameter region for the occurrence of on-resonance bistability, identified by the existence of three positive real solutions for $\Delta_a=\Delta_c=0$, is limited to large values of $n_\eta$ by the condition \eqref{eq:05}. In contrast, when the collective cooperativity is larger than the saturation parameter, only a single real solution of the intracavity light power exists in resonance. The condition \eqref{eq:05} for the onset of on-resonance bistability is represented as dotted black lines in Figs.~\ref{fig:population_both}.

For too strong pumping, as seen in Figs.~\ref{fig:FigBistableBifurcation}, the bistability disappears for all values of $\Gamma$, and we get spectra resembling that of an empty cavity (ECU for empty cavity-like bistable center phase). The maximum pump rate for which bistability occurs is derived from \eqref{eq:A58} in the Appendix \ref{sec:AppThreesols}) as,\Eqn
    {eq:09}{n_\eta \lesssim \frac{N^2g^2}{8\kappa^2}~.}
which is plotted as continuous black horizontal lines. This upper boundary of the bistable region is met when two out of three solutions become degenerate. The analytically derived phase boundaries agree very well with the simulated phase diagrams.

At pump rates below the limit imposed by \eqref{eq:09}, the spectra present either on-resonance bistability (NMBC for normal mode bistable center phase), off-resonance bistability (NMBW for normal mode bistable wing phase), both separately (NMBWC for normal mode bistable wing and center phase) or a merging of both bistable spectral regions (ECBC for empty cavity-like bistable center phase), as we will show below. For the onset of off-resonance bistability close to the normal mode splitting resonances Gripp et al.~\cite{Gripp97b} derived the condition,\Eqn
    {eq:10}{n_\eta > \frac{4\sqrt{N}}{3\sqrt{3}} \frac{(\kappa+\Gamma/2)^3}{g\kappa^2}~,}
shown as dash-dotted black lines. We note that for a collective coupling strength comparable to the cavity decay rate, as in our experimental situation and illustrated in Fig.~\ref{fig:phase_diagram_both}(b), this limit is higher than the critical pump rate according to the condition \eqref{eq:09}, and thus no off-resonance bistability can occur, which leaves the NMBC bistable phase as the only option. For larger collective coupling strength as shown in Fig.~\ref{fig:phase_diagram_both}(a), the orange area delimits a NMBW phase where only off-resonance bistability exists (as shown in Fig.~\ref{fig:phases}(c)), and the green-shaded area delimits a region where both bistabilites coexist in a NMBWC phase; an example of such spectra is also shown in Fig.~\ref{fig:phases}(h). In the region below both conditions \eqref{eq:05} and \eqref{eq:10}, only a unistable solution exists, which characterizes a normal mode unistable (NMU) phase.

When further increasing the pump rate, the bistable regions of the spectrum merge into a ECBC phase. The lower bound of this region is described by the empirically guessed expression,\Eqn
    {eq:11}{n_\eta \gtrsim \frac{N(\kappa+2\Gamma)}{2\kappa}~,}
identified as a dashed black line.

The atomic saturation is an important observable to characterize the different phases of the atom-cavity system. Although  more difficult to measure experimentally than the cavity transmission, it gives further insight on the difference between the on-resonance and off-resonance bistability. Figs.~\ref{fig:population_both}(a) and (b) show the maximum atomic excited state population achieved when scanning the laser frequency, for the same parameters as for Figs.~\ref{fig:phase_diagram_both}(a) and (b). We learn from the graphs that, while the presence of off-resonance bistability does not require full saturation of the atomic population [see for example the region in Fig.~\ref{fig:population_both}(a) that corresponds to the orange region in Fig.~\ref{fig:phase_diagram_both}(a)], it is on the contrary required for the emergence of on-resonance bistability [yellow areas in Figs.~\ref{fig:population_both}(a,b) and corresponding areas in \ref{fig:phase_diagram_both}(a,b)]. This further confirms that the underlying mechanism for this bistability is the appearance of a transparency window caused by full saturation of the atoms. Moreover, the atomic population presents a discontinuity when going from lower to higher pump rates, as the threshold for the on-resonance bistability is crossed. This is also evident from the two insets of Figs.~\ref{fig:population_both}(a) and (b), calculated for the parameters of \cite{Gripp97} and for our experiment, respectively. Since it relies on the full saturation of a quantum two-level transition, the on-resonance bistability cannot stem only from a nonlinear susceptibility and is fully quantum in its origin, at odds with the off-resonance bistability.

\subsection{Comparison to other experiments}

The reason why bistability could be observed in the resonantly coupled regime ($|\Delta_\mathrm{c}|\ll\kappa$) in our experiment is that we are deep in the bad-cavity limit. As seen from Fig.~\ref{fig:FigBistableBifurcation}, bistability around $\Delta_\mathrm{c}=0$ appears over wide ranges of pump rates $\eta_+$ only if $\Gamma\ll\kappa$.

Previous experiments \cite{Gibbs76b,Lambrecht95,Gupta07,Ritter09} carried out with alkali atoms were all performed in the good-cavity limit, due to the large spontaneous decay width $\Gamma$ of the atomic transition. Another experiment carried out by Gothe et al.~\cite{Gothe19} employed ytterbium atoms coupled to a cavity via a narrow transition. However, the finesse of the cavity was so high that the atomic decay rate was still twice the cavity decay rate. Consequently, on resonance, spontaneous emission remained the main decay channel. In order to avoid spontaneous emission leading to heating and trap loss they coupled the atoms dispersively to their cavity, so that the bistable features appeared far from the atomic resonance.

\section{Conclusion}

We have observed and characterized a new mechanism for the emergence of saturation-induced bistability in an optical cavity strongly coupled to $N$ atoms. This new mechanism relies on the complete saturation of a narrow atomic transition quasi-resonant with a cavity mode, where the atomic cloud becomes transparent to the incoming light above a threshold intensity, and a third peak appears in the center between the normal mode resonances. At odds with the previously observed saturation-induced bistability, this new mechanism does not induce an intensity-dependent shift of the frequency for the maximum transmission, and symmetrically induces a bistability window on both sides of the maximum transmission peak. It needs, moreover, a quantum two-level transition to occur, and cannot simply be mimicked by a nonlinear susceptibility in a classical system.

The reason why this new bistability mechanism was just now discovered is that atom-cavity systems are often studied in the good-cavity regime, while this resonant bistability emerges in the bad cavity limit, that is, when $\Gamma<\kappa$. In this range of parameters, the coherence of the light-atom interaction is mainly stored within the atomic coherence. This feature has recently received much attention, due to the possibility of using ultra-narrow transitions of strontium atoms coupled to a resonant optical cavity to build ultra-narrow lasers for metrology \cite{Meiser09,Norcia16c}, decoupled from the cavity noise. The proposed regime for achieving ultra-narrow laser emission relies on strong collective coupling between the atoms and the cavity, as well as on the inversion of the atomic population. Thus, bistable effects such as the one reported here are likely to play an important role.

Motivated by the discovery reported here, and in order to provide a complete picture of the whole phase diagram valid in all regimes, we included in our discussion previous experiments performed in the good-cavity limit. We could comprehensively link all phases to the different relevant parameter ratios, $\Gamma/\kappa$, $\Upsilon_N=Ng^2/\Gamma\kappa$, $s_1=8g^2/\Gamma^2$, and $n_\eta=s_1\eta_+$. This work will guide the exploration of nonlinear coupling between atoms and an optical cavity through up-to-now unexplored regimes, in particular for narrow atomic transitions in the bad cavity limit, where the quantum nature of the atoms cannot anymore be neglected.

The intensity filtering provided by the threshold behavior of the resonant transmission of the cavity can generate non-trivial correlations on the reflected or transmitted light or entanglement between the optical and atomic degrees of freedom. We remember that squeezing has been observed in the light reflected from a 'bad cavity' interacting with atoms \cite{Lambrecht96}. Those interesting prospects will be investigated in future work.

\section*{Acknowledgements}

D.R.~acknowledges the Coordena\c{c}\~ao de Aperfei\c{c}oamento de Pessoal de N\'ivel Superior (CAPES) - Finance Code 001 for scholarships. R.C.T.~acknowledges funding from the Brazilian state agency FAPESP, grant no.~2018/23873-3. Ph.W.C. received support from the project CAPES-COFECUB (Ph879-17/CAPES 88887.130197/2017-01) and FAPESP, grant no.~2022/00261-8. S.S. acknowledges funding by the Deutsche Forschungsgemeinschaft (DFG, German Research Foundation) - 422447846 and 465199066, within Research Unit FOR 5413 "Long-range interacting quantum spin systems out of equilibrium: Experiment, Theory and Mathematics. We thank Igor Lesanovsky for helpful discussions.

\bibliography{Publi2023_RingcavityModes}

\vspace{1cm}
\begin{widetext}
\section*{Appendix} \label{sec:App}
\end{widetext}

\subsection{Open Dicke model for atoms distributed over the mode volume of a ring cavity} \label{sec:AppDicke}

The starting point for the theoretical description is the open system Dicke Hamiltonian for $N$ atoms interacting with two counter-propagating modes of an optical ring cavity. The rotating wave approximation having been made, it reads,\Eqn
	{eq:A01}{\hat H = \hat H_{field}+\hat H_{pump}
	    +\sum_{j=1}^N(\hat H_{atom}^{(j)}+\hat H_{atom:field}^{(j)})~,}
with\Aln
    {eq:A02}{\hat H_{atom}^{(j)} & = -\Delta_\mathrm{a}\hat\sigma_j^+\hat\sigma_j^-
        = -\tfrac{\Delta_\mathrm{a}}{2}(\mathbb{I}_2+\hat\sigma_j^z)\\
    \hat H_{field} & = -\sum\nolimits_\pm\Delta_\mathrm{c}\hat a_\pm^\dagger\hat a_\pm\nonumber\\
    \hat H_{pump} & = -\sum\nolimits_\pm\imath\eta_\pm(\hat a_\pm-\hat a_\pm^\dagger)\nonumber\\
    \hat H_{atom:field}^{(j)} & = g\sum\nolimits_\pm(\hat a_\pm\hat\sigma_j^+e^{\pm\imath kz_j}
        +\hat a_\pm^\dagger\hat\sigma_j^-e^{\mp\imath kz_j})~.\nonumber}
Here, $\Delta_\mathrm{a}\equiv\omega-\omega_\mathrm{a}$ with $\omega_\mathrm{a}=2\pi c/\lambda_\mathrm{a}$ being the atomic resonance frequency, and $\Delta_\mathrm{c}\equiv\omega-\omega_\mathrm{c}$ with $\omega_\mathrm{c}=2\pi m\delta_{fsr}$ being a cavity resonance frequency. $\delta_{fsr}$ is the cavity's free spectral range and $m\in\mathbb{N}$. $\eta_\pm$ are the pump rates of the two counter-propagating modes, $g$ the atom-field coupling strength (corresponding to half the single-photon Rabi frequency), and $g_N=g\sqrt{N}$ is the collective coupling strength. $\hat\sigma_j^+$, $\hat\sigma_j^-$, $\hat\sigma_j^z$ are the usual Pauli matrices applied to the $j$-th atom, so that\Eqn
    {eq:A03}{\mathbf{\hat s} = \tfrac{1}{2}\left(\begin{matrix} \hat\sigma^-+\hat\sigma^+ \\ \imath(\hat\sigma^--\hat\sigma^+) \\
        [\hat\sigma^+,\hat\sigma^-] \end{matrix}\right)}
is the atomic spin operator. $\hat a_\pm$, $\hat a_\pm^\dagger$ are the usual photonic field operators for the counter-propagating modes.

Note that we do not treat photonic recoil $e^{\mp\imath kz_j}$ as a degree of freedom, but just as a parameter depending on the location of the atoms. Decay processes can be considered in a master or in Heisenberg equations via jump operators, $\hat L=\hat\sigma^-$, $\hat a_-$, describing decay processes occurring, respectively, at rates $\gamma=\Gamma,~2\kappa$.

\subsection{Derivation of the equations of motion} \label{sec:AppDerivation}

Using the Hamiltonian \eqref{eq:A01} and for the decay processes $\hat L$ the Lindblad superoperator expression $\mathcal{L}_{\gamma,\hat L}^\dagger\hat A=\gamma(2\hat L^\dagger\hat A\hat L-\hat L^\dagger\hat L\hat A-\hat A\hat L^\dagger\hat L)$, we derive the equations of motion for the individual atomic operators,\Aln
    {eq:A04}{\dot{\hat\sigma}_i^- & = -\imath[\hat\sigma_i^-,\hat H]+\mathcal{L}_{\Gamma/2,\hat\sigma_i^-}^\dagger\hat\sigma_i^-\\
        & = (\imath\Delta_\mathrm{a}-\tfrac{\Gamma}{2})\hat\sigma_i^--\imath g(e^{\imath kz_i}\hat a_+
            +e^{-\imath kz_i}\hat a_-)\hat\sigma_i^z~,\nonumber}
and\Aln
    {eq:A05}{\dot{\hat\sigma}_i^z & = -\imath[\hat\sigma_i^z,\hat H]
            +\mathcal{L}_{\Gamma/2,\hat\sigma_i^-}^\dagger\hat\sigma_i^z\\
        & = 2\imath g(e^{\imath kz_i}\hat a_++e^{-\imath kz_i}\hat a_-)\hat\sigma_i^+\nonumber\\
        & \qquad -2\imath g(e^{-\imath kz_i}\hat a_+^\dagger+e^{\imath kz_i}\hat a_-^\dagger)\hat\sigma_i^-
        -\Gamma(\mathbb{I}_2+\hat\sigma_i^z)\nonumber}
and for the field operators,\Aln
    {eq:A06}{\dot{\hat a}_\pm & = -\imath[\hat a_\pm,\hat H]+\mathcal{L}_{\kappa,\hat a}^\dagger\hat a_\pm\\
        & = (\imath\Delta_\mathrm{c}-\kappa)\hat a_\pm
            -\imath g\sum_j\hat\sigma_j^-e^{\mp\imath kz_j}+\eta_\pm~.\nonumber}

\subsection{Stationary solution} \label{sec:AppSolution}

The stationary solution follows from the expectation values of the equations \eqref{eq:A04} to \eqref{eq:A06},\begin{widetext}
\Aln
    {eq:A11}{\text{(i)} \qquad 0 & = (\imath\Delta_\mathrm{a}-\tfrac{\Gamma}{2})\<\hat\sigma_i^-\>-\imath g(e^{\imath kz_i}
        \<\hat a_+\hat\sigma_i^z\>+e^{-\imath kz_i}\<\hat a_-\hat\sigma_i^z\>)\\
    \text{(ii)} \qquad 0 & = 2\imath g(e^{\imath kz_i}\<\hat a_+\hat\sigma_i^+\>
        +e^{-\imath kz_i}\<\hat a_-\hat\sigma_i^+\>)-2\imath g(e^{-\imath kz_i}\<\hat a_+^\dagger\hat\sigma_i^-\>
        +e^{\imath kz_i}\<\hat a_-^\dagger\hat\sigma_i^-\>)-\Gamma(\mathbb{I}_2+\<\hat\sigma_i^z\>)\nonumber\\
    \text{(iii)} \qquad 0 & = (\imath\Delta_\mathrm{c}-\kappa)\<\hat a_\pm\>-\imath g\sum_je^{\mp\imath kz_j}\<\hat\sigma_j^-\>
        +\eta_\pm\nonumber~.}
\end{widetext}
Neglecting all correlations by setting, e.g.~$\<\hat a_\pm\hat\sigma_i^z\>=\<\hat a_\pm\>\<\hat \sigma_i^z\>$, and analogously for all operator products, we loose phenomena rooted in entanglement, such as spin-squeezing. Using the abbreviations,\Eqn
    {eq:A12}{U_\gamma \equiv U_0-\imath\gamma_0 \equiv \frac{g^2(\Delta_\mathrm{a}-\imath\tfrac{\Gamma}{2})}{\Delta_\mathrm{a}^2+\tfrac{\Gamma^2}{4}} ~~~~~\text{and}~~~~~ \Delta_\kappa \equiv \Delta_\mathrm{c}+\imath\kappa~,}
where $U_0$ has the meaning of a single-photon light shift and $\gamma_0$ of single-photon scattering rate, Eq.~\eqref{eq:A11}(i) resolved by $\<\hat\sigma_i^-\>$ becomes,\Eqn
    {eq:A13}{\<\hat\sigma_i^-\> = \tfrac{U_\gamma}{g}\left(e^{\imath kz_i}\alpha _++e^{-\imath kz_i}\alpha_-\right)\<\hat\sigma_i^z\>~.}
Substituting $\<\hat\sigma_i^\pm\>$ in \eqref{eq:A11}(ii),\Eqn
    {eq:A14}{\left(1+2\left|\tfrac{U_\gamma}{g}\right|^2\left|e^{\imath kz_i}\alpha_++e^{-\imath kz_i}\alpha_-\right|^2\right)\<\hat\sigma_i^z\> = -1~,}
and in (iii),\Eqn
    {eq:A15}{\Delta_\kappa\alpha_\pm-\sum_jU_\gamma\left(e^{\imath kz_j\mp\imath kz_j}\alpha_++e^{-\imath kz_j\mp\imath kz_j}\alpha_-\right)\<\hat\sigma_j^z\> = \imath\eta_\pm~.}
Eliminating $\<\hat\sigma_j^z\>$ from the expressions \eqref{eq:A14} and \eqref{eq:A15},\Eqn
    {eq:A16}{\boxed{\sum_j\frac{U_\gamma(\alpha_\pm+e^{\mp 2\imath kz_j}\alpha_\mp)}
        {1+2|U_\gamma/g|^2|e^{\imath kz_j}\alpha_++e^{-\imath kz_j}
        \alpha _-|^2} = \imath\eta_\pm-\Delta_\kappa\alpha_\pm}~,}
which is the expression \eqref{eq:02} given in the main text.

\subsection{Saturation-induced bistability} \label{sec:AppDisorder}

It is obvious that the steady-state expression \eqref{eq:A16} will generate bistable behavior. With the aim of deriving an analytic solution accounting for our experimental situation we assume that the atomic cloud is homogeneously distributed over the cavity's mode volume. The other extreme case that the atoms are perfectly bunched, $b\equiv\tfrac{1}{N}\sum_je^{2\imath kz_j}=1$, which is the case when the atoms are arranged in an optical lattice, can be treated analogously and produces similar results except from the fact that the mode $\alpha_-$ counter-propagating to the pumped mode $\alpha_+$ may receive a considerable number of photons due to Bragg-enhanced backscattering.

In the case of a homogeneous cloud (no bunching $b=0$) and with one-sided probing, $\eta_-=0$, we may neglect backscattered light altogether, $\alpha_-=0$. Re-scaling via $\bar\Delta_\mathrm{c}\equiv\Delta_\mathrm{c}/\kappa$ and $\bar\Delta_\mathrm{a}\equiv 2\Delta_\mathrm{a}/\Gamma$ and using the definitions of the single-atom cooperativity and the single-photon saturation parameter,\Eqn
    {eq:A21}{\Upsilon \equiv \frac{\Omega_1^2}{\Gamma\kappa} = \frac{4g^2}{\Gamma\kappa} 
        ~~~~~\text{and}~~~~~
        s_1 \equiv \frac{2\Omega_1^2}{\Gamma^2} = \frac{8g^2}{\Gamma^2}~,}
the expression~\eqref{eq:A16} simplifies to,\Eqn
    {eq:A22}{\alpha_+ = \frac{\eta_+}{\kappa}\frac{1}{1-\imath\bar\Delta_\mathrm{c}+\frac{N\Upsilon
        (1+\imath\bar\Delta_\mathrm{a})}{2\left(1+s_1n+\bar\Delta_\mathrm{a}^2\right)}}~,}
so that for the intracavity mean photon number we get,\Eqn
    {eq:A23}{\boxed{n  = \frac{\eta_+^2/\kappa^2}{\left(1+\frac{N\Upsilon/2}
        {1+s_1n+\bar\Delta_\mathrm{a}^2}\right)^2+\left(\bar\Delta_\mathrm{c}-\frac{\bar\Delta_\mathrm{a}N\Upsilon/2}
        {1+s_1n+\bar\Delta_\mathrm{a}^2}\right)^2}}~.}
The 'good-cavity' regime, $s_1<\Upsilon$, and the 'bad-cavity' regime, $s_1>\Upsilon$, are delimited by the ratio between the single-photon saturation and the single-atom cooperativity.\\

The expression \eqref{eq:A23} can be written in terms of a third order polynomial,\Eqn
    {eq:A24}{0 = An^3+Bn^2+Cn+D~,}
with the coefficients,\begin{widetext}\Eqn
    {eq:A25}{\boxed{\begin{array}{rcl}
            A & = & s_1^2(1+\bar\Delta_\mathrm{c}^2) \vspace{.2cm}\\
            B & = & 2s_1\left(\bar\Delta_\mathrm{a}^2+\tfrac{\Upsilon_N}{2}\right)+2s_1\bar\Delta_\mathrm{c}
                \left[\bar\Delta_\mathrm{c}(1+\bar\Delta_\mathrm{a}^2)-\tfrac{N\Upsilon}{2}\bar\Delta_\mathrm{a}\right]
                -s_1s_\eta \vspace{.2cm}\\
            C & = & \left(\bar\Delta_\mathrm{a}^2+\tfrac{\Upsilon_N}{2}\right)^2
                +\left[\bar\Delta_\mathrm{c}(1+\bar\Delta_\mathrm{a}^2)-\tfrac{N\Upsilon}{2}\bar\Delta_\mathrm{a}\right]^2
                -2s_\eta(1+\bar\Delta_\mathrm{a}^2) \vspace{.2cm}\\
            D & = & -n_\eta(1+\bar\Delta_\mathrm{a}^2)^2
        \end{array}}~,}
where we defined the collective cooperativity, the maximum photon number, and the maximum saturation parameter,\Eqn
    {eq:A26}{\Upsilon_N \equiv N\Upsilon+2 ~~~~~\text{and}~~~~~
        n_\eta \equiv \frac{\eta_+^2}{\kappa^2} ~~~~~\text{and}~~~~~
        s_\eta \equiv s_1n_\eta~.}
The roots of the cubic equation are given by,\Eqn
    {eq:A27}{R \equiv \root{3}\of{36CBA-108DA^2-8B^3
        +12\sqrt{3}A\sqrt{4C^3A-C^2B^2-18CBAD+27D^2A^2+4DB^3}}}
\end{widetext}
and\Eqn
    {eq:A28}{X_\pm \equiv \frac{R}{6A}\pm\frac{6AC-2B^2}{3AR}}
so that,\Eqn
    {eq:A29}{n_1 = X_--\frac{B}{3A} ~~~,~~~ 
        n_{2,3} = -\frac{1}{2}X_--\frac{B}{3A}\pm\frac{\imath\sqrt{3}}{2}X_+~,}
from which we obtain the roots $n=|\alpha_+|^2$ and the field amplitude from \eqref{eq:A22}.

The atomic state can be obtained from Eqs.~\eqref{eq:A14} and \eqref{eq:A13},\Aln
     {eq:A30}{\<\hat\sigma_i^z\> & = - \frac{1}{1+2|U_\gamma/g|^2|\alpha_++\alpha_-|^2}\\
        \<\hat\sigma_i^-\> & = \frac{U_\gamma}{g}(\alpha_++\alpha_-)\<\hat\sigma_i^z\>~.\nonumber}

All numerical curves shown in this work are obtained by evaluating the solutions \eqref{eq:A29} discarding all solutions with photon numbers $n\notin\mathbb{R}^+$. Figs.~\ref{fig:FigBistableBifurcation} and \ref{fig:FigBistableRegime} exhibit spectral ranges characterized by the presence of several solutions, while the theoretical spectra of Figs.~\ref{fig:FigAvoidedCrossing}(d-f) and \ref{fig:FigIntensityScan}(c) only show the solution corresponding to the highest photon number.

\subsection{Critical pump rate on resonance} \label{sec:AppResonance}

Let us now discuss the resonant case, $\Delta_\mathrm{c}=0=\Delta_\mathrm{a}$, for which Eq.~\eqref{eq:A22} becomes,\Eqn
    {eq:A33}{\alpha_+ \simeq \frac{\eta_+}{\kappa}\frac{ns_1+1}
        {\frac{1}{2}N\Upsilon+ns_1+1}~.}
From this formula we can easily evaluate the relative importance of collective cooperativity and saturation. Since $N\Upsilon\gg 1$, we get below saturation,\Eqn
    {eq:A34}{\alpha_+ \overset{n\rightarrow 0}{\longrightarrow}
        \frac{\eta_+}{\kappa}\frac{2}{N\Upsilon}~,}
that is, the excitation of the cavity field is heavily suppressed by the normal mode splitting. On the other hand, at high saturation,\Eqn
    {eq:A35}{\alpha_+ \overset{n\rightarrow\infty}{\longrightarrow}
        \frac{\eta_+}{\kappa}~,}
the cavity becomes fully transparent.

For intermediate saturation we evaluate the coefficients \eqref{eq:A25} of the cubic equation (decorated by an apostrophe),\Aln
    {eq:A41}{A' & = s_1^2\\
        B' & = s_1(\Upsilon_N-s_\eta)\nonumber\\
        C' & = \tfrac{1}{4}\Upsilon_N^2-2s_\eta\nonumber\\
        D' & = -n_\eta~.\nonumber}
The cubic equation \eqref{eq:A24} with the resonant coefficients \eqref{eq:A41} has three degenerate real solutions if for some coefficient $n_\mathrm{t}$,\Eqn
    {eq:A42}{n^3+\tfrac{B'}{A'}n^2+\tfrac{C'}{A'}n+\tfrac{D'}{A'} = (n-n_\mathrm{t})^3~.}
In other words,\Eqn
    {eq:A43}{\tfrac{B'}{A'} = -3n_\mathrm{t} ~~~,~~~ 
        \tfrac{C'}{A'} = 3n_\mathrm{t}^2 ~~~,~~~ 
        \tfrac{D'}{A'} = -n_\mathrm{t}^3~.}
From \eqref{eq:A43} we derive,\Aln
    {eq:A45}{3s_1n_\mathrm{t} & = s_\eta-\Upsilon_N\\
        \tfrac{3}{4}\Upsilon_N^2-6s_\eta & = (s_\eta-\Upsilon_N)^2\nonumber\\
        27s_\eta & = (s_\eta-\Upsilon_N)^3~.\nonumber}
The equations are solved for\Eqn
    {eq:A46}{(s_\eta,\Upsilon_N,s_1n_\mathrm{t}) = (0,0,0), (-1,2,-3), (27,18,3)~,}
the only non-trivial physical solution being the last one. This means, that the onset of bistability occurs when,\Eqn
    {eq:A47}{N\Upsilon = 16 ~~~~~\text{and}~~~~~ 
        n_\eta = 27/s_1 ~~~~~\text{and}~~~~~ 
        n_\mathrm{t} = 3/s_1~.}
The onset is entirely governed by an interplay between the saturation parameter $s_\eta$, tuned via the rate at which the cavity is pumped, and the collective cooperativity $\Upsilon_N$, which can be controlled via the number of atoms fed into the cavity mode volume. For our experiment this means, that we observe bistability if more that $N>1232$ atoms interact with the cavity and the pump rate is high enough to inject $n_\eta>2.29$ photons resonantly into the cavity in the absence of atoms. Under these circumstances, we actually find $n_\mathrm{t}>0.25$ photons in the cavity in the presence of $1232$ atoms.

\subsection{Condition for three positive real solutions} \label{sec:AppThreesols}
	
The condition $0=\frac{d}{dn}(A'n^3+B'n^2+C'n+D')$ yields the positions of the turning points of the cubic curve parameterized by \eqref{eq:A24},\Eqn
    {eq:A50}{n_\mathrm{turn} = -\frac{B'}{3A'}\pm\frac{B'}{3A'}\sqrt{1-\frac{3A'C'}{B'^2}}~.}
In order for the cubic curve to have three (not necessarily degenerate) real roots, the cubic curve must have two distinct maxima. That is, \eqref{eq:A50} must yield two real solutions (called turning points), which requires $B'^2<3A'C'$. In order for the cubic curve to have three \textit{positive} real roots, both turning points must be positive. Since $A'>0$ we necessarily need $B'<0$, which leads to,\Eqn
    {eq:A51}{\boxed{\Upsilon_N < s_1n_\eta}~.}
Furthermore, we must satisfy,\Eqn
    {eq:A52}{0 < 1-\frac{3A'C'}{B'^2} < 1~.}
Inserting the coefficients \eqref{eq:A41} we obtain,\Eqn
    {eq:A53}{s_\eta^2-2\Upsilon_Ns_\eta+6s_\eta+\tfrac{1}{4}\Upsilon _N^2 
        > 0 > 6s_\eta-\tfrac{3}{4}\Upsilon_N^2~.}
For large collective cooperativity, $\Upsilon_N\gg 1$, we may simplify,\Eqn
    {eq:A54}{\boxed{\Upsilon_N(1+\tfrac{\sqrt{3}}{2}) < s_1n_\eta < \tfrac{\Upsilon_N^2}{8}}~.}
The criterion \eqref{eq:A54} clarifies, that an essential condition for resonant bistability is a sufficiently large collective cooperativity $\Upsilon_N\gg 8+4\sqrt{3}$. If this is satisfied, one can always find a pump rate $\eta_+$ such that bistability can be reached.
\begin{figure}[ht]
    \centerline{\includegraphics[width=6.7truecm]{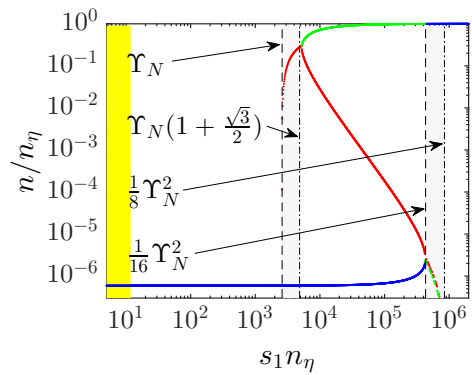}}
    \caption{\small{Numerical simulations of the normalized intracavity photon number as a function of the normalized pump rate for $g_N=1.2\kappa$. They confirm the existence of a single solution for $s_1n_\eta<2600$ [according to \eqref{eq:A51}] or for $s_1n_\eta>845000$ [according to the right-hand inequality of \eqref{eq:A54}]. Bistability with three solutions is observed for the interval specified by the left-hand inequality of \eqref{eq:A54} and by \eqref{eq:A58}, that is, $4850<s_1n_\eta<432000$. Finally, bistability with two solutions is seen in the remaining intervals. The yellow area marks low saturation.}}
    \label{fig:FigBistableRegime}
\end{figure}

The theoretical curve shown in Fig.~\ref{fig:FigBistableRegime} shows the photon numbers calculated in resonance, $\Delta_\mathrm{a}=0=\Delta_\mathrm{ca}$ as a function of pump strengths. Bistable phases are recognized by the presence of several solutions marked in different colors. The transitions between different phases are delimited by the conditions \eqref{eq:A51} and \eqref{eq:A54} and are marked in Fig.~\ref{fig:FigBistableRegime} as vertical black dashed or dash-dotted lines.

\bigskip

In order to derive an analytic expression for the boundary near $s_1n_\eta=432000$ (vertical black dashed line in Fig.~\ref{fig:FigBistableRegime}) we note that the region above the boundary is characterized by the existence of three solutions, two of which are degenerate. These solutions can be found by a procedure similarly to the one previously implemented in Eq.~\eqref{eq:A42}. We plug the coefficients \eqref{eq:A41} into the ansatz,\Eqn
    {eq:A55}{n^3+\tfrac{B'}{A'}n^2+\tfrac{C'}{A'}n+\tfrac{D'}{A'} = (n-n_\mathrm{d})^2(n-n_\mathrm{s})}
and compare the coefficients. This yields a set of equations in the photon numbers $n_\mathrm{s}$, $n_\mathrm{d}$, and $n_\eta$,\Aln
    {eq:A56}{0 & = s_1n_\mathrm{s}+2s_1n_\mathrm{d}+\Upsilon_N-s_1n_\eta\\
    0 & = 8s_1^2n_\mathrm{d}n_s+4s_1^2n_\mathrm{d}^2-\Upsilon_N^2+8s_1n_\eta\nonumber\\
    0 & = -s_1^3n_\mathrm{d}^2n_\mathrm{s}+s_1n_\eta~.\nonumber}

To find an approximate analytic solution of the set of equations \eqref{eq:A56}, we assume, $\Upsilon_N\gg 1$ and $n_\mathrm{s}\gg n_\mathrm{d}$, which simplifies the set to,\Aln
    {eq:A57}{0 & = s_1n_\mathrm{s}-s_1n_\eta\\
    0 & = 8s_1^2n_\mathrm{d}n_s-\Upsilon_N^2+8s_1n_\eta\nonumber\\
    0 & = -s_1^3n_\mathrm{d}^2n_\mathrm{s}+s_1n_\eta~.\nonumber}
The solution is,\Eqn
    {eq:A58}{\boxed{s_1n_\eta \simeq \tfrac{\Upsilon_N^2}{16}}~.}

\subsection{Spectral width of the bistable regime} \label{sec:AppWidth}

Here, we want to calculate the spectral width of the bistable regime, that is, the width of the ridge connecting the normal modes of Fig.~\ref{fig:FigAvoidedCrossing} and the width of the central feature of Fig.~\ref{fig:FigIntensityScan}. Unfortunately, the condition $\Delta_\mathrm{ca}=\Delta_\mathrm{c}-\Delta_\mathrm{a}=0$ entails cumbersome formulas. For simplicity we use instead $\bar\Delta_\mathrm{c}=0$, which corresponds to a cut along a diagonal emphasized by dotted lines in Figs.~\ref{fig:FigAvoidedCrossing}. Assuming a small spontaneous decay rate, $|\bar\Delta_\mathrm{a}|\gg 1$ and a large collective cooperativity, $\Upsilon_N\gg 1$, the coefficients~\eqref{eq:A25} become (decorated by a double apostrophe),\Aln
    {eq:A61}{A'' & = s_1^2\\
       B'' & \simeq s_1(2\bar\Delta_\mathrm{a}^2+\Upsilon_N-s_\eta)\nonumber\\
       C'' & \simeq \bar\Delta_\mathrm{a}^4+\left(\tfrac{\Upsilon_N^2}{4}-2s_\eta\right)\bar\Delta_\mathrm{a}^2\nonumber\\
       D'' & \simeq -n_\eta\bar\Delta_\mathrm{a}^4~.\nonumber}
The coefficient $B''$ determines the slope of the cubic curve in its turning point. For $A''>0$ it must be negative in order to allow for more than one real solution. Thus, we can use the condition $B''=0$ to deduce that, for small enough spontaneous decay rate, the spectral width of the bistable regime is proportional to the Rabi frequency,\Eqn
    {eq:A62}{|\Delta_\mathrm{a}| = \tfrac{\Gamma}{4}\sqrt{s_\eta-\Upsilon_N}
        = \tfrac{1}{2}\sqrt{\tfrac{\Omega_\eta^2}{2}-\tfrac{\Gamma^2\Upsilon_N}{4}}
        \simeq \tfrac{\Omega_\eta}{2\sqrt{2}}~.}

\end{document}